\documentclass[prl,amssymb,amsmath,reprint,superscriptaddress,showpacs]{revtex4-1}
\usepackage{bm}
\usepackage{graphicx}
\usepackage{color}

\usepackage{color}
\definecolor{darkgreen}{rgb}{0.1,0.6,0.1}
\newcommand{\revisionHashi}[1]{#1}
\newcommand{\revisionNEW}[1]{#1}
\newcommand{\revision}[1]{#1}
\newcommand{\revisionR}[1]{#1}
\begin{document}

\title{
Robust Nodal Structure of Landau Level Wave Functions Revealed by Fourier Transform Scanning Tunneling Spectroscopy}

\author{K. Hashimoto}
\affiliation{
Department of Physics, Tohoku University, Sendai 980-8578, Japan}
\affiliation{
JST, ERATO Nuclear Spin Electronics Project, Sendai 980-8578, Japan}

\author{T. Champel}
\affiliation{Universit\'e Joseph Fourier Grenoble I / CNRS UMR 5493,
Laboratoire de Physique et Mod\'elisation des Milieux Condens\'es, B.P. 166,
38042 Grenoble, France}

\author{S. Florens}
\affiliation{Institut N\'{e}el, CNRS and Universit\'{e} Joseph Fourier, B.P.
166, 25 Avenue des Martyrs, 38042 Grenoble Cedex 9, France}

\author{C. Sohrmann}
\affiliation{
Department of Physics and Centre for Scientific Computing, University of Warwick,
Gibbet Hill Road, Coventry CV4 7AL, UK}

\author{J. Wiebe}
\affiliation{
Institute of Applied Physics, Hamburg University, Jungiusstrasse 11, D-20355 Hamburg, Germany}

\author{Y. Hirayama}
\affiliation{
Department of Physics, Tohoku University, Sendai 980-8578, Japan}
\affiliation{
JST, ERATO Nuclear Spin Electronics Project, Sendai 980-8578, Japan}

\author{R. A. R\"{o}mer}
\affiliation{
Department of Physics and Centre for Scientific Computing, University of Warwick,
Gibbet Hill Road, Coventry CV4 7AL, UK}

\author{R. Wiesendanger}
\affiliation{
Institute of Applied Physics, Hamburg University, Jungiusstrasse 11, D-20355 Hamburg, Germany}

\author{M. Morgenstern}
\affiliation{
II. Institute of Physics B and JARA-FIT, RWTH Aachen University, Aachen, D-52074, Germany}
\date{\today }

\begin{abstract}
Scanning tunneling spectroscopy is used to study the real-space local density of states (LDOS)
of a two-dimensional electron system in magnetic field, in particular within higher Landau levels (LL).
By Fourier transforming the LDOS, we find a set of $n$ radial minima at fixed momenta
for the $n$th LL. The momenta of the minima depend only on the inverse magnetic length.
By comparison with analytical theory and numerical simulations, we attribute the minima to the nodes of
the quantum cyclotron orbits, which decouple in Fourier representation
from the random guiding center motion due to the disorder.
\revision{Adequate Fourier filtering reveals the nodal structure in real space in some areas of the sample with relatively smooth
potential disorder.}
%
\end{abstract}

\maketitle


Directly mapping the wave functions of electrons gives the most pertinent access to
the quantum mechanical properties of matter~\cite{Eigler, Manoharan, Manoharan2, Yazdani, Maltezo}. Under a perpendicular
magnetic field $B$ in two-dimensional electron systems (2DES), self-interference
of the circular electronic orbits leads to a standing wave pattern of
probability density, as first calculated by Landau~\cite{textbook}.
The kinetic energy becomes quantized into discrete Landau levels (LL)
$E_n=\hbar \omega_c \left(n+\frac{1}{2} \right)$, with $\omega_c$ the cyclotron
frequency, $\hbar$ Planck's constant, and $n=0,1, 2\ldots$ characterizing the
number of nodes in the LL wave functions.
Experimental observation of this nodal structure has until now remained elusive. The 2DES are usually deeply buried in semiconducting heterostructures, which prevents the use of high resolution scanning tunneling spectroscopy (STS).
Recently, the advent of surface 2DES in doped
semiconductors~\cite{Morg2003b,Hash2008,Becker2010},
graphene~\cite{Niim2009,Mill2010} and on the surface of
topological insulators~\cite{Top1,Top2} has in principle opened the way to such direct high resolution
measurements. This should allow probing the internal structure of the LL wave functions.

However, the Landau energy levels are highly degenerate, so that the
associated wave functions will be strongly disturbed by any perturbation such as
disorder. One has, thus, to deal with the inherent complexity of disorder in
spectroscopic measurements.
More fundamentally, disorder is crucial
for the understanding
of universal quantized Hall conductance~\cite{VonK1980} in 2DES. At high magnetic field, disorder essentially lifts
the LL degeneracy keeping $n$ as a good quantum number due to the large
cyclotron gap.
The electronic motion is then largely decomposed into independent fast cyclotron orbit
and slow drifting motion of the guiding center along equipotential lines of the
smooth disorder landscape.
Most of the equipotential lines in the bulk being closed, this picture provides a
simple localization mechanism that ensures a reservoir of localized electronics states
between \revisionHashi{LLs},
 ultimately responsible for the formation of wide
quantized plateaus of Hall conductance~\cite{textbook}.
This semiclassical picture was recently confirmed by STS real space imaging
of the electronic probability density~\cite{Hash2008,Mill2010}.
The latter was found in the lowest \revisionHashi{LL},
 LL0, to follow equipotential lines with
a transverse spread on the scale of the magnetic length $l_B=\sqrt{\hbar/|e|B}$.
Within the higher \revisionHashi{LLs},
 LL$n$ for $n>0$, the drift motion is expected
to be accompanied by larger cyclotron orbits characterized by the quantum Larmor
radii $R_n=l_B \sqrt{2n+1}$. This larger spread of the meandering LL wave functions in
the transverse direction to the guiding center motion should contain signatures of the
LL nodal structure for $n\geq1$.
Note that these nodes can be related to phase singularities of LL wave functions with $n$ a winding number \cite{Czerwinski,Champel2007}.  This topological origin confers some robustness to the nodal structure which should be viewed as a key property of quantum Hall states.
The purpose of this Letter is to show how to reveal these interference effects.
\begin{figure*}[tb]
\includegraphics[width=0.9\linewidth]{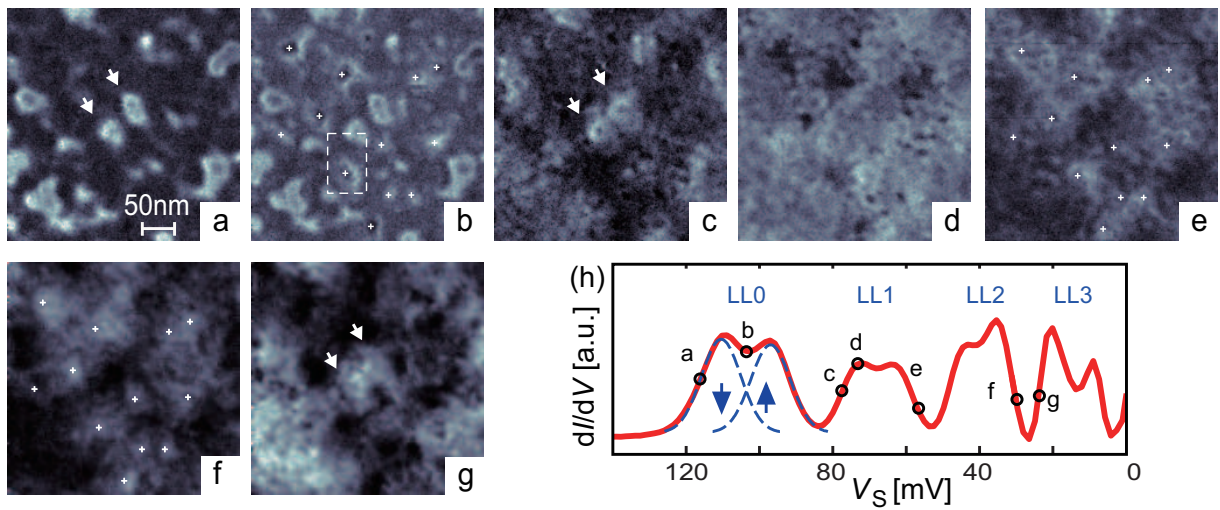}
\caption{(color online) (a)--(g) Real-space LDOS taken at $B=6$ T
for the sample voltages $V_{\rm s}$ marked as circles in (h);
measurements are done by lock-in technique with the modulation voltage $V_{\rm mod} = 1$ mV
after stabilizing the tip at a current $I_{\rm stab} = 0.1$ nA
and a sample voltage $V_{\rm stab} = 150$ mV;
all d$I/$d$V$ images are recorded in the same area and are displayed
using the same color scale; values of the chosen sample voltage are:
$V_s = -117$ mV (a), -103 mV (b), -79 mV (c), -73 mV (d), -57 mV (e), -29.4 mV (f),  -23.1 mV (g).
White arrows at the same positions in (a), (c), (g)
mark localized states, which exhibit additional nodal structure in (c), (g).
Crosses in (b), (e), and (f) mark states localized at potential hills;
dashed rectangle \revisionHashi{in (b) marks an area around a drift trajectory}.
(h) Spatially averaged d$I/$d$V$ curve using the 1600 curves recorded in the area
of (350 nm)$^2$; 
dashed lines: Gaussian fits to the two spin levels of LL0.}
\label{Fig1}
\end{figure*}

The quantum states in disordered LLs are delicate to analyze \revision{in detail},
because drift and cyclotron motion are, strictly speaking, not disentangled.
We will show that a useful spectroscopic analysis can still be made
by two-dimensional (2D) Fourier transform of the STS data, which allows to deconvolute
the contribution from the discrete LL orbits and from the disorder induced drift motion.
The nodal structure of LL$n$ is revealed by a succession of $n$ {\it ring-like patterns}
in momentum-space, given by a set of fixed and disorder-independent minima
within the momentum scale $R_n/l_B^2$. Analytical and numerical calculations are performed
to vindicate our findings.
Moreover, a methodology to show the nodal structure in real-space is demonstrated.

The 2DES was prepared by 1\% monolayer Cs adsorption
on an n-type InSb(110) surface~\cite{Bett2001} with donor (acceptor) density
$N_{\rm D}=9\cdot 10^{21}$ m$^{-3}$ ($N_{\rm A}=5\cdot 10^{21}$ m$^{-3}$)~\cite{Hash2008}.
The STS measurements~\cite{Hash2008} were performed in ultra-high vacuum at $B = 6$ T and
temperature $T = 0.3$ K \cite{Wieb2004}.
Figures~\ref{Fig1}(a)--(g) show images of the differential conductivity d$I/$d$V$
recorded within the same spatial area for seven different sample voltages
$V_s$ spanning the four spin-split LLs from LL0 to LL3 [see the spatially averaged
d$I$/d$V$ curve shown in Fig.~\ref{Fig1}(h)].
These spatial maps represent the local density of states (LDOS) consisting of
all wave functions at energy $E\simeq eV_s$ within the experimental energy resolution
of 2.5 meV \cite{Hash2008,Morg2003a}, \revisionNEW{see discussion in
\cite{noteSupInfo}}.
In the lower tail of LL0 [Fig.~\ref{Fig1}(a)], several spatially isolated closed loops
are visible, which correspond to individual localized states encircling
potential minima as shown in Ref.~\cite{Hash2008}. The same states are visible within
the lower tail of the higher spin level of LL0 [Fig.~\ref{Fig1}(b)],
however, superimposed to the states marked by white crosses
which belong to the upper tail of the lower spin level and encircle potential maxima.
Similar patterns are found at the same positions in the lower tail of higher LLs
[marked by arrows for LL1 in Fig.~\ref{Fig1}(c) and LL3 in Fig.~\ref{Fig1}(g)],
but they are spatially wider and are, thus, more difficult to discriminate. \revision{The} widening
\revision {of the drift states marks}
the progressive increase of $R_n$ \revision{with $n$.
Careful inspection of the localized states in the LL's lower tail
[Figs.~\ref{Fig1}(a),(c),(g), see arrows] and in the upper tail
[Figs.~\ref{Fig1}(b),(e),(f)] reveals that the closed loops exhibit more complex
oscillatory patterns perpendicular to the loop in higher LLs.}

\begin{figure*}
\includegraphics[width=1.0\linewidth]{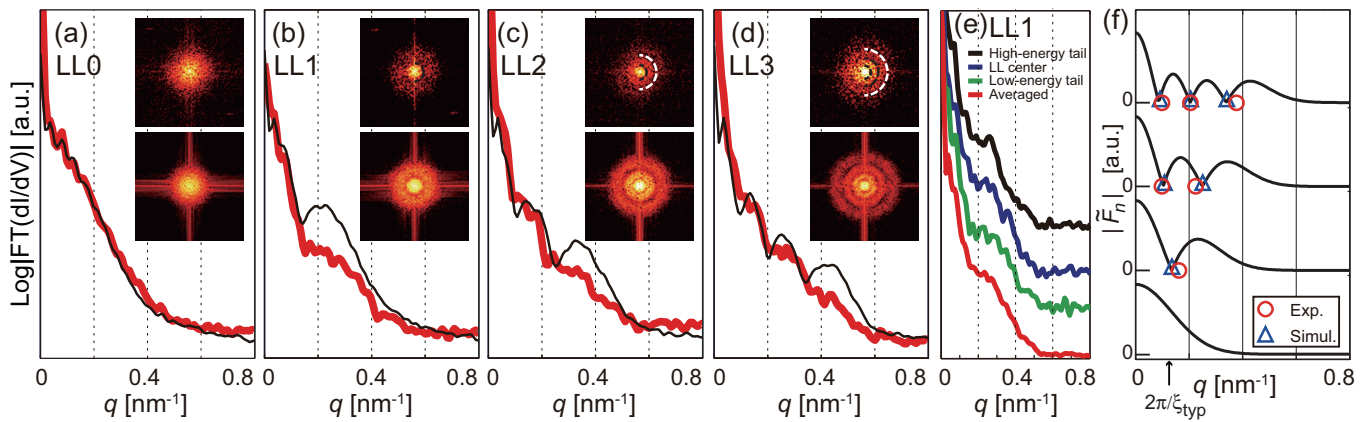}
\caption{(color) (a)--(d) Logarithmic absolute values of angular-averaged
Fourier transformation (FT) of LDOS obtained in the tail of the LLs;
thick red lines: experimental data taken from LDOS of Fig.~\ref{Fig1}(a) (LL0),
Fig.~\ref{Fig1}(c) (LL1), Fig.~\ref{Fig1}(f) (LL2), Fig.~\ref{Fig1}(g) (LL3);
voltage averaging: 2 mV; thin black lines: FT of numerical simulations for an energy
in the tail of the Landau band, $E=\hbar \omega_c\cdot (n-0.4+1/2)$;
the curves are stretched along the $y$ axis in order to match at ${\bf q}=0$ nm$^{-1}$;
insets: FT-LDOS; upper panels: experiment with dashed half circles
indicating ${\bf q}$ of simulated LDOS minima, lower panels: simulations;
(e) Angular-averaged FT-LDOS of LL1 with (red curve) and without voltage averaging
as derived from Fig.~\ref{Fig1}(e) (black), Fig.~\ref{Fig1}(d) (blue),
Fig.~\ref{Fig1}(c) (green); (f) Comparison of absolute Fourier-transformed structure factor
$|\tilde{F}_n({\bf q})|$ (shifted vertically for clarity) as given 
in Eq.~(\ref{FnFourier}) (black lines) and discrete mimima
of the curves in (a)--(d); red circles: experiment; blue triangles: simulation; black arrow marks
$|\bf q|$ $= {2 \pi}/{\xi_{\rm typ}}$ $= 0.126$ nm$^{-1}$
with $\xi_{\rm typ} = 50$ nm.}
\label{Fig2}
\end{figure*}
In order to relate the oscillatory features to the nodal structure of
the LL wave functions, we firstly overcome the difficulty
that the cyclotron motion is randomly correlated to the guiding center motion.
\revisionNEW{This partly masks the nodal structure within our real-space data.  
Note that LL mixing would also blur this structure. However, in the present
experimental conditions (large cyclotron gap and smooth disorder, see
Ref.~\cite{Hash2008}), LL mixing plays a negligible role (see discussion in
\cite{noteSupInfo}). In this case,}
previous works~\cite{Raikh1995,Champel2009,Champel2011} have demonstrated
an {\it exact} decomposition of the LDOS as a function of position ${\bf r}$ and energy $E$:
\begin{eqnarray}
\rho({\bf r},E) =  \int\!\!\!
\frac{d^{2}{\bf R}}{2 \pi l_{B}^{2}}
\sum_{n=0}^{+ \infty}
F_{n}({\bf R}-{\bf r}) A_n({\bf R},E).
\label{ldos}
\end{eqnarray}
Here, we neglect, for the sake of simplicity, thermal smearing effects and
the experimental energy resolution, which could be implemented straightforwardly.
$A_n({\bf R},E)$ corresponds to the guiding center spectral density in LL$n$,
which encodes the complicated dynamics induced by the disorder.
The key object in Eq.~(\ref{ldos}) is the so-called structure factor
\begin{eqnarray}
F_{n}({\bf R}) = \frac{(-1)^{n}}{\pi l_{B}^{2}}
L_{n}\left(\frac{2 {\bf R}^{2}}{l_{B}^{2}} \right) e^{-{\bf R}^{2}/l_{B}^{2}},
\label{Fn}
\end{eqnarray}
where $L_{n}(z)$ is the Laguerre polynomial of degree $n$ containing $n$
oscillations \revision{(or, equivalently, nodes)}.

The structure factor $F_n({\bf R})$ \revision{is not strictly positive and,
while associated with the quantum cyclotron motion in LL$n$,
cannot be interpreted directly as the wave function probability density.}
Instead, $F_n({\bf R})$ corresponds to a Wigner distribution~\cite{Champel2009},
because the physical real-space of the guiding center coordinates ${\bf R}=(X,Y)$
is in fact associated to a pair of quantum conjugate variables~\cite{Goerbig},
owing to the commutation relation $[\hat{X},\hat{Y}]=il_B^2$. The convolution
given by Eq.~(\ref{ldos}) illustrates the difficulty in resolving in real-space
the nodes of the Landau states that are built in the structure factor.
Indeed, the sharp nodal structure of $F_n({\bf R})$ is
not only blurred in $\rho({\bf r},E)$, but the smeared nodal patterns must
also follow the random meanders of the guiding center trajectories encoded in
$A_n({\bf R},E)$. However, by performing a 2D Fourier transformation (FT) of the LDOS,
a simpler product form is achieved
\begin{equation}
\widetilde{\rho}({\bf q},E) =
\sum_{n=0}^{+ \infty} \widetilde{F}_{n}({\bf q})
\widetilde{A}_n({\bf q},E),
\label{ldosFT}
\end{equation}
where the FT of the structure factor reads:
\begin{equation}
\widetilde{F}_{n}({\bf q}) =
L_n\left(\frac{l_B^2 {\bf q}^2}{2}\right) e^{-l_B^2 {\bf q}^2/4}.
\label{FnFourier}
\end{equation}
For a smooth disorder potential characterized by a large correlation length
$\xi\gg l_B$, $\widetilde{A}_n({\bf q},E)$ possesses disorder-induced structures
mainly at short wave vectors $|{\bf q}|\simeq \revision{2\pi}\xi^{-1}$,
and should barely vary
around the larger momentum scale $|{\bf q}|\simeq R_n/l_B^2\propto 1/l_B$.
Thus, distinguishable momentum-variations in the FT of the LDOS~(\ref{ldosFT})
should arise from the {\it universal} structure factor~(\ref{FnFourier}).
We now examine this issue both using our STS data and numerical simulations
of a disordered 2DES at high $B$ field.

In order to reveal the nodal structure of LLs in momentum-space encoded into
$\widetilde{F}_n({\bf q})$, we
proceed with the FT of the LDOS images of
Figs.~\ref{Fig1}(a) (LL0),~\ref{Fig1}(c) (LL1),~\ref{Fig1}(f) (LL2),
and~\ref{Fig1}(g) (LL3).
\revision{Since the FT of the LDOS displays sign changes due both to the
structure factor and to the random spectral density in
momentum space, we focus our discussion on the absolute value of the FT signal.}
We also use a voltage averaging of 2 mV of the data in order to improve contrast.
The upper insets in Figs.~\ref{Fig2}(a)--(d) illustrate the resulting Fourier-transformed
LDOS (FT-LDOS) images. They exhibit a single radial modulation  for LL1, changing into
double (LL2) and triple (LL3) radial modulations at the $|{\bf q}|$ positions marked
by dashed half circles.
\revision{In order to smoothen the random contributions of the spectral density,
we perform an angular average of the signal
as shown by the curves (thick red lines) in Figs.~\ref{Fig2}(a)--(d).
A consequence of this angular averaging is that deviations from perfect angular symmetry and 
noise within the experiments
will add up to a (possibly $q$-dependent) finite background.
Therefore the expected zeros from the structure factors are shifted
upwards and become minima.}
At LL0 [Fig.~\ref{Fig2}(a)], the resulting FT curve decreases monotonically,
composing a disk structure in the FT-LDOS (upper inset).
However, at LL1 [Fig.~\ref{Fig2}(b)], the FT curve exhibits a dip
at $|{\bf q}| = 0.16$ nm$^{-1}$ followed by a broad hump.
This results in an additional ring-like structure surrounding a smaller disk
in the FT-LDOS image (upper inset). The number of dips/humps in the FT curve,
i.e.\ the number of additional rings in the 2D-image,
increases to 2 at LL2 [Fig.~\ref{Fig2}(c)] and to 3 at LL3 [Fig.~\ref{Fig2}(d)].
This general trend is almost quantitatively reproduced
by numerical Hartree simulations~\cite{Sohr2007,Hash2008}
[thin black lines in Fig.~\ref{Fig2}(a)--(d) and lower insets].
Note that especially the position of the minima in the simulated FT curve
shows good agreement with the position of the dip in the experimental one.
The simulations diagonalize the wave functions at $B=6$ T
within a random disorder potential calculated using $N_{\rm D}$ and $N_{\rm A}$ of
the InSb sample~\cite{Sohr2007,Hash2008}. In order to transform the resulting 3D disorder
potential into 2D, a folding with the confined wave function parallel to $B$
deduced from the triangular well approximation \cite{Ando1982} is used.
The resulting 2D disorder is lower than in the experiment, since it ignores disorder
from the Cs atoms \cite{Hash2008}. This lower disorder explains
the stronger features in the simulated FT, however, without any effect
on the position of the minima. Moreover, the position of the minima is robust within each
disorder-broadened LL \revision{as shown in Fig.~\ref{Fig2}(e). The same dip position is observed
for all voltages,} i.e.\ for localized as well as for extended states.
This is remarkable, since the corresponding real-space LDOS
[Figs.~\ref{Fig1}(c)--(e)] shows very different patterns due to the complicated guiding-center motion.
The FT results confirm experimentally that 
\revisionHashi{$\widetilde{\rho}$ from Eq. \ref{ldosFT}} is dominated by the energy-independent structure factor
$\widetilde{F}_n({\bf q})$ at the large momentum scale $|{\bf q}|\simeq l_B^{-1}$. More precisely,
the minima appearing at larger ${\bf q}$ are direct fingerprints of the distinct nodal structure
within each LL. Fig.~\ref{Fig2}(f) shows a direct comparison of the minima positions
deduced from Fig.~\ref{Fig2}(a)--(d) with $|\tilde{F}_n({\bf q})|$.
The mimima from the FT of experiment (circles) and simulation (triangles)
quantitatively match the minima in $|\tilde{F}_n({\bf q})|$.

\begin{figure}
\includegraphics[width=1\linewidth]{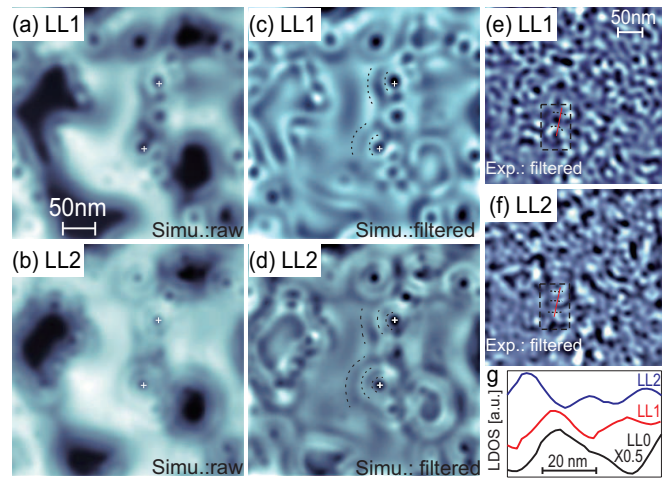}
\caption{(color online) Real-space nodal structure of LLs along the guiding center trajectories;
(a), (b) Raw real-space LDOS from numerical simulations within tail of LL1 (a) and LL2 (b)
[$E=\hbar \omega_c(n-0.4+1/2)$, $n=1,2$].
(c), (d) The same real-space LDOS as in (a),(b) after Fourier high-pass filtering \cite{software}
within \revisionHashi{$|\bf q|$ $\ge$ $0.14$ nm$^{-1}$}.
Crosses in (a)--(d) mark representative potential minima.
(e), (f) Fourier band-pass filtered real-space LDOS from Figs.~\ref{Fig1}(e),(f) 
\revisionHashi{with additional low-pass filtering $|\bf q|$ $\le$ $0.52$ nm$^{-1}$ 
to reduce experimental noise}; 
dashed rectangles mark the same area as in Fig.~\ref{Fig1}(b); 
black dotted lines mark double (LL1) and triple (LL2) lines.
\revisionHashi{(g) L}ine sections along the \revisionHashi{same} line marked \revisionHashi{in (e)(LL1),(f)(LL2) 
and filtered LDOS (LL0)~\cite{noteSupInfo} from Fig. \ref{Fig1}(b) (shifted vertically for clarity).}
}
\label{Fig3}
\end{figure}
We now come back to the real space data, where the nodal structure can be observed by appropriate filtering.
We use the fact that the disorder-induced drift is encoded
up to $|\bf q|$ $\simeq 0.13$ nm$^{-1}$ [black arrow in Fig.~\ref{Fig2}(f)] with our potential disorder
\cite{typicalLength}. 
The key characteristic feature of the real-space nodal structures is encoded at
larger $|\bf q|$.
Thus, by performing band-pass filtering of the FT-LDOS at large  $\bf q$
and a subsequent inverse Fourier transform, we  can  identify in real space
the transverse nodes decoupled from the potential disorder (see Fig.~\ref{Fig3}).
Since the scales of disorder and guiding center motion are not strongly different, adequate borders of the filtering
are essential to improve the visibility of the nodal structure as outlined in the supplement \cite{noteSupInfo}.
\revisionHashi{In the \revisionR{numerical} simulations, we} find faint LDOS corrugations around the guiding center trajectories
encircling the potential minima, marked by crosses in Figs.\ \ref{Fig3}(a) and (b),
while sharp oscillations perpendicular to the drift trajectory, marked by dotted lines, appear
for LL1 (two maxima) and for LL2 (three maxima) after Fourier filtering 
\cite{software} [Figs. \ref{Fig3}(c),(d)].
The double and triple lines are distinct in flat areas of the potential (as marked),
while in other areas where neighboring potential maxima are
close, interference between their corresponding drift states prohibits a clear distinction.
\revisionNEW{Within the experimental data, the nodal line structure is more
difficult to find due to the large number of overlapping potential extrema.
Careful comparison of the data obtained within different \revisionHashi{LLs} in
combination with Fourier filtering
\revisionR{nevertheless reveals some regions} where the
transition from single via double to triple lines \revisionR{is observable}.
\revisionR{For example, consider the} \revisionHashi{ marked lines (in dashed rectangles)
in the filtered LDOS of LL1, LL2 [Figs.~\ref{Fig3}(e),(f)] and LL0~\cite{noteSupInfo}}
[see line sections in Fig.~\ref{Fig3}(g)]}.
%

In summary, FT of spatial STS data leads to
a decoupling of the LDOS structures attributed to the quantum cyclotron orbit
and to the complex disorder-dependent guiding-center motion.
The former exhibits $n$ minima at universal $|\bf q|$ for the $n$th 
\revisionHashi{LL}.
These minima are associated to the $n$ oscillations of the Landau structure factor.
By adequately Fourier filtering the real-space LDOS, the nodal structure is
identified in real space.
Our findings demonstrate that Landau quantization implies disorder independent
universal features on the microscopic scale. This complements the well known universal
quantum Hall plateaus.

\begin{acknowledgments}
K.H. thanks T. Nakajima, M. Koshino, N. Shibata, H. Akera, K. Nomura for helpful
discussions. We acknowledge DFG-projects Wi 1277/15-3, Mo 858/11-1, SFB 508-B4
and JSPS KAKENHI 21740219 for
financial support. Part of the calculations were performed at the U.K.
National Grid Service.
\end{acknowledgments}


\begin{thebibliography}{99}
\bibitem{typicalLength}
This is evaluated by ${2 \pi}/{\xi_{\rm typ}}$ $= 0.126$ nm$^{-1}$
with $\xi_{\rm typ} = 50$ nm being the typical correlation length of our disorder
as determined from the distance of localized states, e.g.\ in Fig.~\ref{Fig1}(a).
\end{thebibliography}
\end{document}